\documentclass[letter]{aa}
\usepackage{graphicx}
\usepackage{bm,natbib,url,wasysym}
\usepackage[varg]{txfonts}
\usepackage{xcolor}
\usepackage{amsmath}
\usepackage[colorlinks,allcolors=blue]{hyperref}

\bibpunct{(}{)}{;}{a}{}{,} 
\graphicspath{{./figs/}{./png/}}

\newcommand{\EQ}{\begin{equation}}
\newcommand{\EN}{\end{equation}}
\newcommand{\EQA}{\begin{eqnarray}}
\newcommand{\ENA}{\end{eqnarray}}

\newcommand{\Sec}[1]{Sect.~\ref{#1}}

\newcommand{\Secs}[2]{Sects.~\ref{#1} and \ref{#2}}

\newcommand{\Fig}[1]{Fig.~\ref{#1}}

{}
{}
{}

{}
{}
{}
{}
{}
{}
{}
{}
{}
{}
{}
{}
{}
{}
{}
{}
{}
{}

{}
{}
{}

{}
{}
{}
{}

{}

{}
{}

\newcommand{\uu}{\mbox{\boldmath $u$} {}}

\newcommand{\AAA}{\mbox{\boldmath $A$} {}}

\newcommand{\s}{\,{\rm s}}

\newcommand{\m}{\,{\rm m}}

\definecolor{upforestgreen}{rgb}{0.0, 0.55, 0.13}

\begin{document}

\titlerunning{Data-driven model of the solar corona above an active region}
\authorrunning{Warnecke \& Peter}

\title{Data-driven model of the solar corona above an active region}
\author{J. Warnecke\inst{1} \and H. Peter\inst{1}}
\institute{Max-Planck-Institut f\"ur Sonnensystemforschung,
  Justus-von-Liebig-Weg 3, D-37077 G\"ottingen, Germany\\
\email{warnecke@mps.mpg.de}\label{inst1} 
}

\date{Received 1 March 2019 / Accepted 27 March 2019}
\abstract{}{%
We aim to reproduce the structure of the corona above a
solar active region as seen in the extreme ultraviolet (EUV) using a
three-dimensional magnetohydrodynamic (3D MHD) model.
}{%
The 3D MHD data-driven model solves the induction equation and the
mass, momentum, and energy balance. To drive the system, we feed the
observed evolution of the magnetic field in the photosphere of the
active region AR\,12139 into the bottom boundary. This creates a hot
corona above the cool photosphere in a self-consistent way. We
synthesize the coronal EUV emission from the densities and
temperatures in the model and compare this to the actual coronal
observations.
}{%
We are able to reproduce the overall appearance and key features of
the corona in this active region on a qualitative level. The model shows long loops, fan
loops, compact loops, and diffuse emission forming at the same
locations and at similar times as in the observation. Furthermore, the
low-intensity contrast of the model loops in EUV matches the observations.
}{%
In our model the energy input into the corona is similar as in the
scenarios of fieldline-braiding or flux-tube tectonics, that is,
energy is transported to the corona through the driving of the
vertical magnetic field by horizontal photospheric motions.
The success of our model shows the central role that this process
plays for the structure, dynamics, and heating of the corona.
}
\keywords{Magnetohydrodynamics (MHD) -- Sun: magnetic fields -- Sun: corona -- methods: numerical
}

\maketitle


\section{Introduction}
\label{sec:intro}

When observed in the extreme ultraviolet (EUV), the corona of the Sun above an active
region is dominated by plasma loops over a range of temperatures from
just below one million to several million Kelvin.
The magnetic field in the corona channels the plasma and guides the
energy flux.
A {one-dimensional (1D)} model can capture the distribution and variation of
intensities and flows along a loop, at least if the variation in energy input is correctly prescribed.
It cannot account for the spatial complexity of the real
corona, however.

A 3D model can account for not only the complex interaction of the
various magnetic features, but also self-consistently provides
the spatial and temporal distribution of the energy input \cite[with
all its limitations;][]{P15}.
The first of these 3D models \citep[][]{GN02,GN05a,GN05b} created a
loop-dominated corona; this confirmed the field-line braiding
\citep[][]{Par88} or  flux-tube tectonics scenarios \citep{PHT02}. 
One main question was then and still is now whether such a model
can recreate the actually observed corona, if the model is driven by the observed
changing magnetic field in the photosphere.

So far, such models have been compared to observations in a generic
sense and with good success.
For example, results have been compared with respect to average
quantities such as emission line Doppler shifts
\cite[][]{PGN04,PGN06,HHDC10}.
Alternatively, individual features in the models have been picked out that
resemble actually observed structures. This provided interesting matches in
terms of the width of coronal loops \cite[][]{PB12} or transient
 UV bursts \cite[][]{HAP17}.

The driving of the magnetic field in the photosphere in these models
is prescribed by a photospheric velocity driver that mimics the solar
granulation \citep{BP11,BP13}. Alternatively, magnetoconvection
models are directly included in the model \citep{GCHHLM2011, Rempel17} or are fed
in through the boundary condition \citep{CPBC14}.
Using a flux emergence model motivated by observations for the
photospheric input, \cite{CRC18} were able to reproduce the emission
signature of a C-class flare.
\cite{BBP13,BBP14} used an observed magnetogram to drive the coronal
evolution, but this was limited to an active region that was just slightly
larger than an X-ray bright point.

Magneto-frictional models have been used to recreate the corona driven
by the changing observed photospheric magnetic field, but by design,
these models do not provide information on temperature and density
\cite[][]{CR12}.
They can instead only derive proxies for the coronal emission
that is expected.
In these models, the currents are furthermore essentially almost
(anti-) parallel to the magnetic field; this assumption is not
fully valid above  active regions \citep{PWCC15,WCBP17}.

Here we use a data-driven 3D magnetohydrodynamic (MHD) model
in which the observed (variable) magnetic field of an active
  region in the photosphere is considered as a lower boundary condition.
Most importantly, the model allows synthesizing the coronal emission.
A direct comparison can therefore be made between the model and the
observed coronal emission from the exact time of the driving
magnetogram.

\begin{figure}[t!]
\begin{center}
\includegraphics[width=\columnwidth]{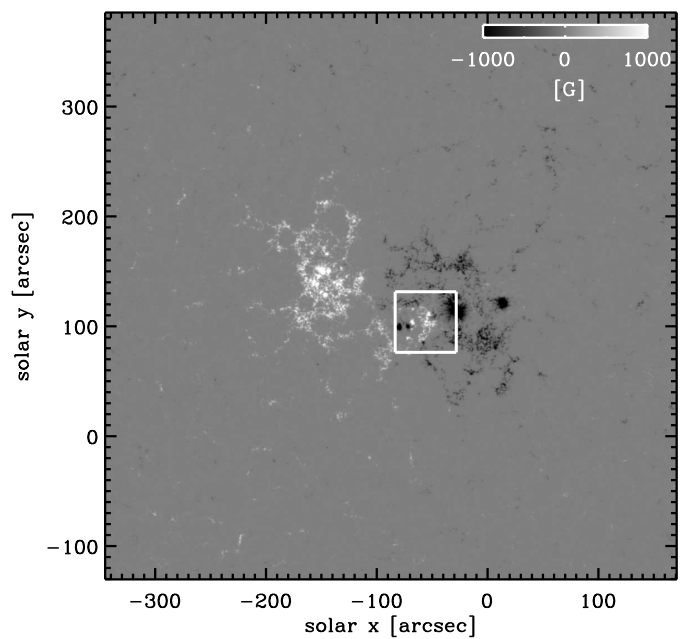}
\end{center}\caption[]{
Observed photospheric magnetic field that drives our simulation. 
We show the line-of-sight magnetogram from 16 August 2014
at 23:30:48 UT of active region AR\, 12139 observed with HMI on board
of SDO.
The temporal evolution of this magnetogram is used as input
for the photospheric magnetic field in the simulation, see \Sec{sec:model}.
The small square indicates a region of interested where a compact loop
is observed, see \Secs{sec:comp}{sec:heat}.
}\label{Blos}
\end{figure}

\begin{figure*}[t!]
\begin{center}
\includegraphics[width=2\columnwidth]{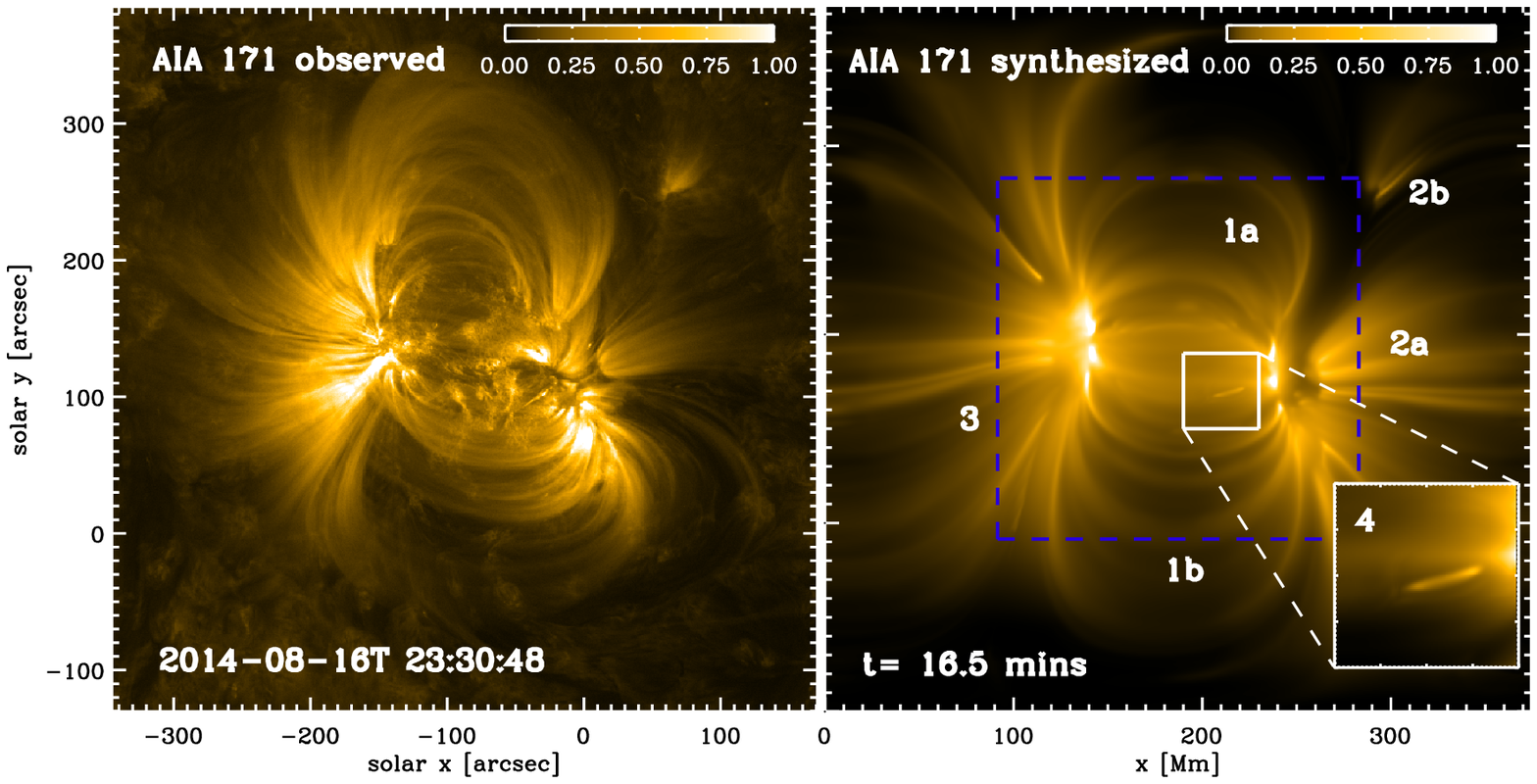}
\end{center}\caption[]{
Comparison of the observed emission and the emission as synthesized from
model. The left panel shows the emission of the AIA 171~\AA\ channel
of AR\, 12139 on 16 August 2014 at 23:30:48 UT near disk center. The
right panel shows the synthesized emission of the same channel from
the simulation as viewed from the top of the computation box. 
For better visibility, we use a non-linear scaling of the images (power of 0.7 for the observation, and power of 0.4 for the synthesized
  emission). The color bars reflect this. The peak count value in
  the observations (corresponding to 1.00) is 3500 DN/pixel;
  this is a factor of six lower for the model. This difference corresponds
  to a factor of 2.5 in density (see \Sec{sec:limitations}).
The inlay shows a zoom-in of the region indicated by the
white square. It shows a compact small loop. The color scaling is
linear here. 
The observations and the model cover the same physical space on the
Sun with a field of view of $(515.9\arcsec)^2$ corresponding to $(374.4\,$Mm$)^2$. 
The numbers indicate the features discussed in
\Sec{sec:comp}. The blue dashed rectangle indicates the zoom-in used in
\Fig{j2rho}.
}\label{obs}
\end{figure*}

\section{Data-driven 3D  MHD model}
\label{sec:model}

We numerically solved the 3D resistive MHD equations, that is,\ the
induction equation together with the mass, momentum, and energy
balance, from the surface of the Sun into the corona.
For this we used the {\sc Pencil
  Code}\footnote{\url{https://github.com/pencil-code/}} with its
special module to account for the physics of the corona
\citep{BP11,BP13}.
This solves for the vector potential $\AAA$, the velocity $\uu$, the
density $\rho,$ and the temperature $T$ in a fully self-consistent and
time-dependent way.

One key element of a coronal model is the inclusion of (Spitzer) heat
conductivity along the magnetic field that depends on temperature as
$T^{5/2}$.
We used non-Fourier heat flux evolution and semi-relativistic Boris correction \cite[e.g.,][]{Boris1970,Rempel17},
both newly implemented into the {\sc Pencil Code} to speed up the
simulation significantly \cite[see][for details]{CP18,WB18}.
The plasma was cooled by optically thin radiative cooling calculated
from a prescribed radiative loss function.
The details of the model are presented in \citet{BP11,BP13} and \citet{WB18}
and are not repeated here.
We used a magnetic diffusivity of $\eta{=}5{\times}10^9\,\m^2\s^{-1}$,
ensuring a mesh Reynolds number of around unity, and a viscosity of 
$\nu{=}10^{10}\,\m^2\s^{-1}$, similar to the Spitzer value at coronal
temperatures and densities.

Our computational domain was a Cartesian box with
$1024\times1024\times512$ grid points, representing
$374\times374\times80$\,Mm$^3$ on the Sun. This is large enough to host a
typical active region.
We used periodic boundary conditions in the horizontal $x$ and $y$
directions.
At the top boundary, the box was closed for all thermodynamic
quantities, and we applied a potential field condition for the magnetic
field. 
At the bottom boundary ($z{=}0$), which represents the solar surface, the
temperature and density were fixed.
Here, we prescribed the photospheric velocities using a granulation
driver that mimics the distribution of flows in time and space
comparable to the observed motions. We followed the original description by
\cite{GN05a} for this.

The central ingredient of our model is the implementation of the
bottom boundary for the magnetic field.
Here, we fed a time series of observed values for the (vertical)
magnetic field and thus drove the evolution of the magnetic field in
the photosphere so that it matched the observed evolution. 
Because the time cadence of the magnetograms is much slower than the
time step of the simulation, we interpolated between the magnetograms
that were closest in time for every time step of the numerical model.
Photospheric velocities also act on the magnetic
  field at the bottom boundary and alter it. To ensure that the magnetic
  field at the bottom boundary continued to evolve according to the observations,
  we employed a relaxation scheme. This smoothly forces the field at the
  bottom boundary to follow as prescribed by observations. We chose a
  timescale of 10 min for this relaxation, motivated by the general
  timescale of the granular magnetic fields. This approach allowed us to generate upward-directed Poynting flux and simultaneously ensured that we remained close to the observed state \citep[see][for details]{BP11,BP13,WB18}.

To feed the simulation, we used a time series of
line-of-sight magnetic field measurements of active region AR\,12139
from the Helioseismic and Magnetic Imager \citep[HMI;][]{HMI} that begin
on 16\ August\ 2014, start at 23:14:53 UT, and have a cadence of 45\,s
(\Fig{Blos}).
The region was very close to disk center, and we therefore used the
line-of-sight magnetic field for the vertical component in our
simulation.
The grid spacing of the model (366\,km) is the same as the plate scale
of the HMI observation (0.5\arcsec per pixel).
We adjusted the edges of the magnetograms to ensure that ethey fulfilled the
horizontally periodic boundary conditions.
We first ran the simulation with the magnetic field from the first
snapshot of the time series to let the temperature and density reach a
quasi-stationary state. This took about four solar hours.
We then began to feed the time series of the changing magnetic field
into the bottom boundary; this drives the evolution in the computational
domain.
The simulation then evolved for another half hour. We
  focus our analysis below on the snapshot at 16.5 min.

\section{Comparison with observations}
\label{sec:comp}

\begin{figure}[t!]
\begin{center}
\includegraphics[width=\columnwidth]{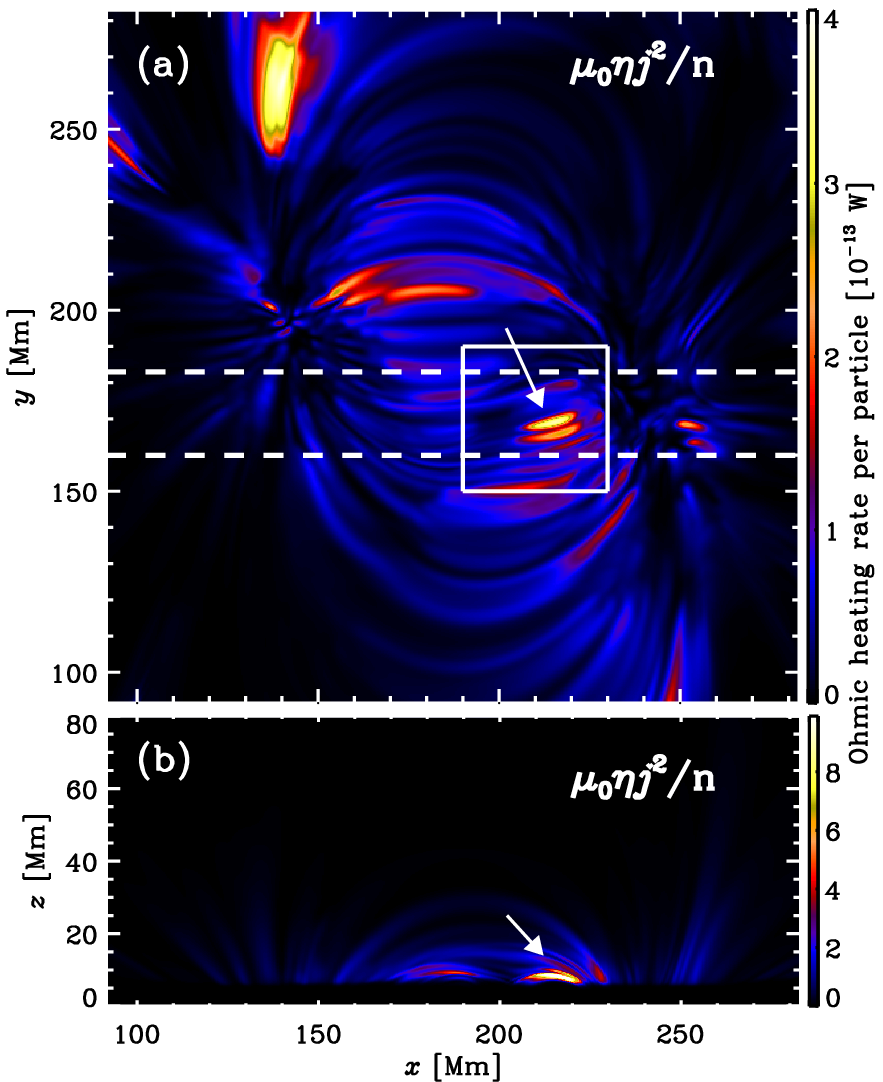}
\end{center}\caption[]{
Energy deposition in terms of Ohmic heating rate per particle $\mu_0\eta j^2/n$. Panel a
shows the top view of an average over 2 min (14.5 to 16.5 min) and in the 
vertical direction from $z=$0 to 30 Mm. Panel b shows an average over
a stripe with $y=$160 to 183 Mm, as indicated in
panel a (also averaged over 2 min). The white square shows the region of
interest with the compact small loop (indicated by arrows in
both panels); compare with \Fig{obs}.
}\label{j2rho}
\end{figure}

The main goal of this study comparing real observations to
the coronal emission as synthesized from our 3D MHD active region model, which is
driven by a time series of actually observed magnetograms.
For this we derived EUV emission as it would be observed by the
Atmospheric Imaging Assembly \cite[AIA;][]{LTAB2012}.
Based on the temperature and density in the model, we used the
temperature response kernel \cite[][]{AIA:2012}\footnote{Implemented
  in SolarSoft (\url{http://www.lmsal.com/solarsoft/}).} of the
171~\AA\ channel. This channel is dominated by emission from \ion{Fe}{x}
that originates from just below 1\,MK.

Overall, the numerical model reproduces the dipole-like structure
of the active region.
In \Fig{obs} we show the comparison of the model as viewed from
straight above (left) and the observations near disk center (right) at
the same time, that is, the model was evolved using the
time-dependent observed magnetograms to the same time as the
observations shown here (16.5 mins).
The peak emission from the observations and
  the simulation differs by a factor lower than six, which corresponds to
  differences in density smaller than a factor of 2.5; see
  \Sec{sec:limitations} for a detailed discussion.
Despite this quantitative difference, we find an overall qualitative agreement, in
particular for the following four features (numbers as in
\Fig{obs}).

{\it (1) Long loops}.
We find that loops with lengths of 100\,Mm to 200\,Mm connect opposite
polarities at the edges of the active region in the northern (top) and
southern areas.
These loops are associated with the large-scale potential-like
magnetic field of the active region.
The model loops appear to be less strongly helical than the observed loops, probably because the
observed magnetograms also lack strong magnetic helicity (see \Sec{sec:limitations}).
Like in the observations, the model shows quite a few distinct long
loops in the southern part. Some even lie at roughly the same
location (1a). The observations show a more
complex broad bundle of loops in the northern part, where the model only shows a single
long loop (1b). 
The long model loop in the north still shows a rather broad
structure with a clear loop in the middle. 

{\it (2) Fan loops}.
In our model we find fan loops, in particular, on the western (right)
side of the active region. They appear at the same locations as very similar
features that are visible in the observations. 
Several thinner structures quickly diverge and form a funnel-type
structure in which the thinner strands are embedded (2a).
Our model even reproduces a smaller  feature of this type outside of
the main part of the active region (2b) in an area of enhanced magnetic field strength (cf.\ \Fig{Blos}).
The fan loops in this setup also appear  because of the horizontal periodic domain; see the discussion
  in \Sec{sec:limitations}.

{\it (3) Diffuse emission}.
Loops in the corona have a rather low contrast, they often stick out of
a diffuse non-resolved background by only 10\% to 30\%
\citep{DZM03,PBKF13}.
We see this general pattern in our model as well.
Most of the thin loops are embedded in much thicker structures of
diffuse emission. 
In our model this is due to the high magnetic resistivity, but it might
well reflect the situation on the real Sun \cite[for a discussion of
the resistivity in MHD models, see][]{P15}.

{\it (4) Compact loops}.
In the core of an active region, observations show an abundance of
small transient features, which may be low-lying loops that are related to
small-scale magnetic patches in the photosphere \cite[][]{PBKF13}.
In the model we see only a few of these, probably because of the
limited spatial resolution (see \Sec{sec:limitations}).
The example shown in the inset of \Fig{obs} only exists for less than
10\,min in the simulation and is indeed a low-lying loop (cf.\
Sect.\,\ref{sec:heat}) that is rooted in two small opposite-polarity patches
that evolve quickly.

\section{Energy deposition in the corona}
\label{sec:heat}

The coronal structures that appear in the model do so because energy is deposited along field lines that are at their footpoints
driven by horizontal motions.
Here, we briefly discuss the relation of the loops that appear to the
energy input per particle.
In \Fig{j2rho} we display the distribution of the energy input at the
same time as the snapshot of the emission shown in \Fig{obs}, but
integrated in time for the 120\,s leading up to that time.
This accounts for the Alfv\'en transit time through the coronal
structure so that disturbances of the magnetic field have time to
spread.

When we integrate the energy input per particle vertically,
loop features become visible that are similar to those seen in emission (compare \Fig{j2rho}
to the emission in the blue dashed rectangle in \Fig{obs}).
We see more spatial variation along the field lines in the energy input
than in coronal intensity.
The reason is that the energy is efficiently redistributed through
(Spitzer) heat conduction parallel to the magnetic field. Heat
conduction quickly evens out temperature inhomogeneities and leads to a
comparably constant temperature along a field line and thus a
comparably constant coronal emission along the loop.
The appearance of the EUV loops is directly linked to the
energy input into the corona \cite[][]{PB12} and is (in our
model) a direct consequence of the energy injection near the loop
footpoints \cite[at the height where plasma-$\beta$ is about unity;
cf.][]{CPBC14}.

While the model evolves, that is, while the changing photospheric field drives
the system, compact coronal loops appear in a transient fashion. One
of these short loops is highlighted by the white squares in all the figures.
It clearly connects opposite magnetic polarities (\Fig{Blos}), and the
photospheric motions lead to the strong heating (per particle) in what
then appears as the short EUV loop (inset of \Fig{obs}).
The greater heating is clearly visible when the computational
box is viewed from the top and from the side (arrows in \Fig{j2rho}). These
small features might be related to the miniature loops that were found in
observations \cite[][]{PBKF13}.
Here the loop is showing up transiently because at low heights
and high densities, radiative cooling apparently is
efficient and thus the lifetime of the structure is short.
The overall large-scale loop structure including the long
  loops, fan loops, and diffusive emission is even visible before
  we evolve the magnetic field at the bottom boundary according to the
  observations. However, the small compact loops only appear when we use a time-dependent
  photospheric field for driving. The time evolution of the magnetic field deposits
  more energy into the corona on smaller scales.

\section{Limitations of the model}
\label{sec:limitations}

This model is generally successful in reproducing the observed
corona of the active region, but several limitations need to
be kept in mind.
The main limitation probably is the moderate spatial resolution that
does not allow us to properly  resolve the small-scale magnetic features
on granular scales of 1000\,km and smaller.
Most of the shortcomings of our model might be traced to this
limitation.

A higher spatial resolution would reveal smaller features of the
photospheric magnetic field that drives the model.
These small magnetic patches would give rise to a higher
(transient) energy input that would cause more small-scale hot
structures at low heights to appear as compact transient loops.
Observationally, there are clear indications that such small-scale opposite
polarities cause heating of coronal structures
\cite[][]{CPSB17,CSRP19}.
With the current setup, we need observations of the photospheric
magnetic field of the whole active region, and such data are available
only at a moderate spatial resolution of about 1\arcsec, such as those provided by
HMI that we use here.
High-resolution observations would provide only an insufficient
field of view that does not cover the whole active region.

Our  model does not produce  regions at high temperature of
5\,MK or above at sufficiently high density to give rise to X-ray loops
in the core of the active region.
Again, this could be due to the lack of resolution, which prevents opposite
polarities from canceling and from providing high-energy fluxes into the corona
\citep[cf.\ observations by][]{CPS18}.
In a model with a smaller computational domain, \cite{AH14} indeed reported that
plasma was heated to flare-like temperatures in compact (few Mm)
structures.
In general, we would therefore expect a higher energy input into the upper
atmosphere if we were able to feed the model with magnetic field
data at significantly higher spatial resolution.

As a consequence of the possibly too low energy flux in the model, the
model density might be too low as well. According to the scaling laws of
\cite{RTV78}, the coronal density scales roughly with the heat input to
the power of 4/7.
The densities in our model, and consequently, the count rates, might therefore be
lower than in the observed active region. 
The count rates in the model are too low by about a factor of six, which means that the density is too low by about a factor of 2.5.
This probably also affects the density distribution along the
loop and is the reason that we scaled our model emission nonlinearly with a
power of 0.4 in \Fig{obs} (while the observations are scaled closer to
linear with a power of 0.7).

The  loops in our model corona look more like those in a potential field than like the
loops in the observations. 
In our model we only prescribed the vertical component of the magnetic field, but not the horizontal component (which is not available at 45\,s
cadence and for this large field of view with HMI).
We would therefore miss any helical component of the magnetic field, which
would lead to a more twisted appearance of
the magnetic field.
Moreover, this would increase the energy input into the system and thus
help alleviate the problem described above, at least to some extent.

Finally, we assume periodic boundary conditions in our model (as do most
other 3D MHD models of the solar atmosphere).
This implies that the field lines of the fan loops on the western
(right) side of the active region leave the box on that side and enter
the box again on the other side. 
In the model, the fan loops therefore connect to an active region (different from the first). On
the real Sun, such a connected active region would be far away, at a distance that is far larger than the size of an active region. 
We tried to account for this in the model by surrounding the active region by enough quiet Sun, so that the appearance of
the fan loops is hopefully realistic.

None of the above limitations are expected to alter the main
conclusion of our study: our data-driven model can account
for the structures seen on the real Sun.
Future models will have to address these remaining problems step by step, however.

\section{Conclusions}

We showed that we can reproduce many aspects of the corona
above an active region using a data-driven 3D MHD model.
The emission that we synthesized from the model qualitatively reproduces the observed
coronal features of this active region:  long loops, fan-like loops,
and small transient loops in the active region core. 
We also showed that these model features are embedded in a  diffuse background of
coronal emission, as is the case on the real Sun, where a typical
loop has a low contrast of only 10\% to 30\% to the background.

Our data-driven model shows that the structure of the observed
photospheric magnetic field and its temporal evolution fully govern
the appearance of the corona in an active region. Driving the magnetic
field at the surface induces currents in the corona at just the correct
places: here the plasma is heated and forms EUV loops in the
model at exactly the same place where they also appear in the real observations.
The energy input in our model is solely based on the driving of the
magnetic field and is thus very similar to the scenarios of field-line
braiding \citep{Par88} and flux-tube tectonics scenarios \citep{PHT02}. 
The success of our model therefore supports these
heating scenarios. 

The main shortcoming of our work is the lack of spatial resolution of
the photospheric field: to cover a full active region, observational
data are available only at a moderate resolution of about 1\arcsec.
However, we can expect smaller magnetic patches to play a significant
role in energizing the corona.
This could lead to a higher energy input, and in particular, also to a
higher structuring of the corona in the core of the active region.
Likewise, if we included the horizontal component of the magnetic field, the energy input would increase, and would in particular lead to a more strongly
twisted appearance of the coronal loops in the model. This would bring them
closer to the actual observations.

Our
data-driven 3D MHD model does reproduce the overall appearance of the
corona in an active region despite the limitations outlined in Sect.\,\ref{sec:limitations}.
This first look at the results shows that the driving of the corona by
the observed magnetic field at the surface level indeed gives rise to
coronal structures that appear in the model at (roughly) the same time
and location as in the observations of the real Sun.

\begin{acknowledgements}
We thank Lakshmi Pradeep Chitta and Sven Bingert for discussion leading to this work.
Simulations have been carried out on supercomputers at
GWDG, on the Max Planck supercomputer at RZG in Garching, in the facilities hosted by the CSC---IT
Center for Science in Espoo, Finland, which are financed by the
Finnish ministry of education. 
J. W.\ acknowledges funding by the Max-Planck/Princeton Center for
Plasma Physics and 
 from the People Programme (Marie Curie
Actions) of the European Union's Seventh Framework Programme
(FP7/2007-2013) under REA grant agreement No.\ 623609.
\end{acknowledgements}

\bibliographystyle{aa}
\bibliography{paper}

\begin{thebibliography}{35}
\expandafter\ifx\csname natexlab\endcsname\relax\def\natexlab#1{#1}\fi

\bibitem[{Archontis \& Hansteen(2014)}]{AH14}
Archontis, V. \& Hansteen, V. 2014, The Astrophysical Journal, 788, L2

\bibitem[{{Bingert} \& {Peter}(2011)}]{BP11}
{Bingert}, S. \& {Peter}, H. 2011, \aap, 530, A112

\bibitem[{{Bingert} \& {Peter}(2013)}]{BP13}
{Bingert}, S. \& {Peter}, H. 2013, \aap, 550, A30

\bibitem[{{Boerner} {et~al.}(2012){Boerner}, {Edwards}, {Lemen}, {Rausch},
  {Schrijver}, {Shine}, {Shing}, {Stern}, {Tarbell}, {Title}, {Wolfson},
  {Soufli}, {Spiller}, {Gullikson}, {McKenzie}, {Windt}, {Golub}, {Podgorski},
  {Testa}, \& {Weber}}]{AIA:2012}
{Boerner}, P., {Edwards}, C., {Lemen}, J., {et~al.} 2012, \solphys, 275, 41

\bibitem[{{Boris}(1970)}]{Boris1970}
{Boris}, J.~P. 1970, NRL Memorandum Report, 2167

\bibitem[{{Bourdin} {et~al.}(2013){Bourdin}, {Bingert}, \& {Peter}}]{BBP13}
{Bourdin}, P.-A., {Bingert}, S., \& {Peter}, H. 2013, \aap, 555, A123

\bibitem[{{Bourdin} {et~al.}(2014){Bourdin}, {Bingert}, \& {Peter}}]{BBP14}
{Bourdin}, P.-A., {Bingert}, S., \& {Peter}, H. 2014, \pasj, 66, S7

\bibitem[{{Chatterjee}(2018)}]{CP18}
{Chatterjee}, P. 2018, Geophys. Astrophys. Fluid Dyn., submitted
  [\eprint[arXiv]{1806.08166}]

\bibitem[{{Chen} {et~al.}(2014){Chen}, {Peter}, {Bingert}, \&
  {Cheung}}]{CPBC14}
{Chen}, F., {Peter}, H., {Bingert}, S., \& {Cheung}, M.~C.~M. 2014, \aap, 564,
  A12

\bibitem[{{Cheung} \& {DeRosa}(2012)}]{CR12}
{Cheung}, M.~C.~M. \& {DeRosa}, M.~L. 2012, \apj, 757, 147

\bibitem[{{Cheung} {et~al.}(2018){Cheung}, {Rempel}, {Chintzoglou}, {Chen},
  {Testa}, {Mart{\'{\i}}nez-Sykora}, {Sainz Dalda}, {DeRosa}, {Malanushenko},
  {Hansteen}, {De Pontieu}, {Carlsson}, {Gudiksen}, \& {McIntosh}}]{CRC18}
{Cheung}, M.~C.~M., {Rempel}, M., {Chintzoglou}, G., {et~al.} 2018, Nature
  Astronomy, 3, 160

\bibitem[{{Chitta} {et~al.}(2018){Chitta}, {Peter}, \& {Solanki}}]{CPS18}
{Chitta}, L.~P., {Peter}, H., \& {Solanki}, S.~K. 2018, \aap, 615, L9

\bibitem[{{Chitta} {et~al.}(2017){Chitta}, {Peter}, {Solanki}, {Barthol},
  {Gandorfer}, {Gizon}, {Hirzberger}, {Riethm{\"u}ller}, {van Noort}, {Blanco
  Rodr{\'{\i}}guez}, {Del Toro Iniesta}, {Orozco Su{\'a}rez}, {Schmidt},
  {Mart{\'{\i}}nez Pillet}, \& {Kn{\"o}lker}}]{CPSB17}
{Chitta}, L.~P., {Peter}, H., {Solanki}, S.~K., {et~al.} 2017, \apjs, 229, 4

\bibitem[{{Chitta} {et~al.}(2019){Chitta}, {Sukarmadji}, {Rouppe van der
  Voort}, \& {Peter}}]{CSRP19}
{Chitta}, L.~P., {Sukarmadji}, A.~R.~C., {Rouppe van der Voort}, L., \&
  {Peter}, H. 2019, \aap, 623, A176

\bibitem[{{Del Zanna} \& {Mason}(2003)}]{DZM03}
{Del Zanna}, G. \& {Mason}, H.~E. 2003, \aap, 406, 1089

\bibitem[{{Gudiksen} {et~al.}(2011){Gudiksen}, {Carlsson}, {Hansteen}, {Hayek},
  {Leenaarts}, \& {Mart{\'{\i}}nez-Sykora}}]{GCHHLM2011}
{Gudiksen}, B.~V., {Carlsson}, M., {Hansteen}, V.~H., {et~al.} 2011, \aap, 531,
  A154

\bibitem[{{Gudiksen} \& {Nordlund}(2002)}]{GN02}
{Gudiksen}, B.~V. \& {Nordlund}, {\AA}. 2002, \apjl, 572, L113

\bibitem[{{Gudiksen} \& {Nordlund}(2005{\natexlab{a}})}]{GN05b}
{Gudiksen}, B.~V. \& {Nordlund}, {\AA}. 2005{\natexlab{a}}, \apj, 618, 1031

\bibitem[{{Gudiksen} \& {Nordlund}(2005{\natexlab{b}})}]{GN05a}
{Gudiksen}, B.~V. \& {Nordlund}, {\AA}. 2005{\natexlab{b}}, \apj, 618, 1020

\bibitem[{{Hansteen} {et~al.}(2017){Hansteen}, {Archontis}, {Pereira},
  {Carlsson}, {Rouppe van der Voort}, \& {Leenaarts}}]{HAP17}
{Hansteen}, V.~H., {Archontis}, V., {Pereira}, T.~M.~D., {et~al.} 2017, \apj,
  839, 22

\bibitem[{{Hansteen} {et~al.}(2010){Hansteen}, {Hara}, {De Pontieu}, \&
  {Carlsson}}]{HHDC10}
{Hansteen}, V.~H., {Hara}, H., {De Pontieu}, B., \& {Carlsson}, M. 2010, \apj,
  718, 1070

\bibitem[{{Lemen} {et~al.}(2012){Lemen}, {Title}, {Akin}, {Boerner}, {Chou},
  {Drake}, {Duncan}, {Edwards}, {Friedlaender}, {Heyman}, {Hurlburt}, {Katz},
  {Kushner}, {Levay}, {Lindgren}, {Mathur}, {McFeaters}, {Mitchell}, {Rehse},
  {Schrijver}, {Springer}, {Stern}, {Tarbell}, {Wuelser}, {Wolfson}, {Yanari},
  {Bookbinder}, {Cheimets}, {Caldwell}, {Deluca}, {Gates}, {Golub}, {Park},
  {Podgorski}, {Bush}, {Scherrer}, {Gummin}, {Smith}, {Auker}, {Jerram},
  {Pool}, {Soufli}, {Windt}, {Beardsley}, {Clapp}, {Lang}, \&
  {Waltham}}]{LTAB2012}
{Lemen}, J.~R., {Title}, A.~M., {Akin}, D.~J., {et~al.} 2012, \solphys, 275, 17

\bibitem[{{Parker}(1988)}]{Par88}
{Parker}, E.~N. 1988, \apj, 330, 474

\bibitem[{{Peter}(2015)}]{P15}
{Peter}, H. 2015, Philosophical Transactions of the Royal Society of London
  Series A, 373, 20150055

\bibitem[{{Peter} \& {Bingert}(2012)}]{PB12}
{Peter}, H. \& {Bingert}, S. 2012, \aap, 548, A1

\bibitem[{{Peter} {et~al.}(2013){Peter}, {Bingert}, {Klimchuk}, {de Forest},
  {Cirtain}, {Golub}, {Winebarger}, {Kobayashi}, \& {Korreck}}]{PBKF13}
{Peter}, H., {Bingert}, S., {Klimchuk}, J.~A., {et~al.} 2013, \aap, 556, A104

\bibitem[{{Peter} {et~al.}(2004){Peter}, {Gudiksen}, \& {Nordlund}}]{PGN04}
{Peter}, H., {Gudiksen}, B.~V., \& {Nordlund}, {\AA}. 2004, \apjl, 617, L85

\bibitem[{{Peter} {et~al.}(2006){Peter}, {Gudiksen}, \& {Nordlund}}]{PGN06}
{Peter}, H., {Gudiksen}, B.~V., \& {Nordlund}, {\AA}. 2006, \apj, 638, 1086

\bibitem[{{Peter} {et~al.}(2015){Peter}, {Warnecke}, {Chitta}, \&
  {Cameron}}]{PWCC15}
{Peter}, H., {Warnecke}, J., {Chitta}, L.~P., \& {Cameron}, R.~H. 2015, \aap,
  584, A68

\bibitem[{{Priest} {et~al.}(2002){Priest}, {Heyvaerts}, \& {Title}}]{PHT02}
{Priest}, E.~R., {Heyvaerts}, J.~F., \& {Title}, A.~M. 2002, \apj, 576, 533

\bibitem[{{Rempel}(2017)}]{Rempel17}
{Rempel}, M. 2017, \apj, 834, 10

\bibitem[{{Rosner} {et~al.}(1978){Rosner}, {Tucker}, \& {Vaiana}}]{RTV78}
{Rosner}, R., {Tucker}, W.~H., \& {Vaiana}, G.~S. 1978, \apj, 220, 643

\bibitem[{{Schou} {et~al.}(2012){Schou}, {Scherrer}, {Bush}, {Wachter},
  {Couvidat}, {Rabello-Soares}, {Bogart}, {Hoeksema}, {Liu}, {Duvall}, {Akin},
  {Allard}, {Miles}, {Rairden}, {Shine}, {Tarbell}, {Title}, {Wolfson},
  {Elmore}, {Norton}, \& {Tomczyk}}]{HMI}
{Schou}, J., {Scherrer}, P.~H., {Bush}, R.~I., {et~al.} 2012, \solphys, 275,
  229

\bibitem[{{Warnecke} \& {Bingert}(2018)}]{WB18}
{Warnecke}, J. \& {Bingert}, S. 2018, Geophys. Astrophys. Fluid Dyn., submitted
  [\eprint[arXiv]{1811.01572}]

\bibitem[{{Warnecke} {et~al.}(2017){Warnecke}, {Chen}, {Bingert}, \&
  {Peter}}]{WCBP17}
{Warnecke}, J., {Chen}, F., {Bingert}, S., \& {Peter}, H. 2017, \aap, 607, A53

\end{thebibliography}

\end{document}